\documentclass[twocolumn]{article}

\usepackage[big]{dgruyter}
\usepackage{longtable}
\usepackage{pdflscape}
\usepackage{translations}
\usepackage{romannum}
\usepackage{graphicx}	
\usepackage{amsmath}	
\usepackage{amssymb}	
\usepackage{multirow}
\usepackage{bm}
\usepackage{authblk}

\usepackage{color}
\usepackage{caption}
\setcitestyle{authoryear,open={(},close={)}}

\newcommand\kms{\ensuremath{\mbox{km}\,\mbox{s}^{-1}}}
\newcommand\Teff{\ensuremath{T_\mathrm{eff}}}
\newcommand\logg{\ensuremath{\log g}}
\newcommand\vsini{\ensuremath{v_{e}\sin i}}
\newcommand\micro{\ensuremath{\xi_{\mathrm{t}}}}

\newcommand\met{\ensuremath{[M/H]}}

\begin{document}

\author*[1]{Marwan Gebran$^1$}
\author[2]{Ian Bentley$^2$}
\author[3]{Rose Brienza$^1$}
\author[4]{Fr\'ed\'eric Paletou$^3$}

\runningauthor{M. Gebran}  
\affil[1]{Department of Chemistry and Physics, Saint Mary’s College, Notre Dame, IN 46556, USA, E-mail: mgebrane@saintmarys.edu}

\affil[2]{Department of Physics, Florida Polytechnic University, Lakeland, FL, 33805}

\affil[3]{Universit\'e de Toulouse, Observatoire Midi--Pyr\'en\'es, Irap, Cnrs, Cnes,  14 av. E. Belin, F--31400 Toulouse, France}

\baretabulars 

\articletype{Research Article}

\title{Deep Learning application for stellar parameters determination: \\
III- Denoising Procedure}
\runningtitle{DL for stellar parameters III}

\begin{abstract}
{In this third paper in a series, we investigate the need of spectra denoising for the derivation of stellar parameters. We have used two distinct datasets for this work. The first one contains spectra in the range  of 4,450-5,400 \AA\ at a resolution of 42,000 and the second in the range of 8,400--8,800 \AA\ at a resolution of 11,500. We constructed two denoising techniques, an autoencoder, and a Principal Component Analysis. Using random Gaussian noise added to synthetic spectra, we have trained a Neural Network to derive the stellar parameters \Teff, \logg, \vsini, \micro, and \met\ of the denoised spectra. We find that, independently of the denoising technique, the stellar parameters accuracy values do not improve once we denoise the synthetic spectra. This is true with and without applying data augmentation to the stellar parameters Neural Network.} 
\end{abstract}


 \journalname{Open Astronomy}
\DOI{DOI}
  \startpage{1}
  \received{..}
  \revised{..}
  \accepted{..}

  \journalyear{2024}
  \journalvolume{1}

\maketitle

\section{Introduction}
Observations in astronomy have always been associated with noise. Trying to minimize the noise is one of the needs of astronomers. Several observation techniques have been suggested to reduce the noise in spectra, however, once the observation is performed, the only way to proceed is to apply mathematical algorithms that can improve the Signal-to-Noise Ratio (SNR) of the data. These techniques involve but are not limited to Gaussian smoothing \citep{2020arXiv200709539C}, median filtering \citep{9083712}, wavelet denoising \citep{wavelet}, and Principal Component Analysis \citep{BACCHELLI2006606,ZHANG20101531,dedde,8531438}. More recently, and with the advancement of computational power, Deep Learning algorithms started to be used in that purpose. \cite{2022MNRAS.509..990G} used a Convolutional Denoising autoencoder to decrease the noise of synthetic images of state-of-the-art radio telescopes like LOFAR \citep{2013A&A...549A..11O}, MeerKAT \citep{5109671}, and MWA \citep{2013PASA...30....7T}. The technique was applied on different kinds of corrupted input images. The autoencoder was able to effectively denoise images identifying and extracting faint objects at the limits of the instrumental sensitivity. The authors state that their autoencoder was capable of removing noise while preserving the properties of the regions of the sources with SNR as low as 1.  \cite{2023MNRAS.526.3037S} used a variational autoencoder to denoise optical SDSS spectra of galaxies (\citealt{2000AJ....120.1579Y}).Their main goal was to denoise the spectra while keeping the important information they can retrieve from low SNR galaxy spectra and avoiding the use of sample averaging methods (smoothing or spectral stacking). They tested the method in the context of large optical spectroscopy surveys by simulating a population of spectra with noise to mimic the ones at galaxies at a redshift of $z=0.1$. Their results showed that the technique can recover the shape and scatter of the mass-metallicity relation in this sample.

In this work, we introduce two types of spectral denoising techniques, autoencoders \citep{1863696.1863746,Baldi2011AutoencodersUL} and Principal Component Analysis (PCA, \citealt{WOLD198737,MACKIEWICZ1993303}). We test the need of the denoising technique on the derived stellar parameters: effective temperature \Teff, surface gravity \logg, equatorial projected rotational velocity \vsini, microturbulence velocity \micro, and the overall metallicity \met. These stellar parameters are derived using the Neural Network introduced our previous work \citep{2022OAst...31...38G,2023OAst...32..209G,2024Astro...3....1G}. The paper is divided as follows: Sec.~\ref{database} introduces the calculation of both datasets and noisy spectra, Sec.~\ref{AE} explains the autoencoder construction used in the denoising procedure, and Sec.~\ref{PCA} describes the denoising technique using Principal Component Analysis.  Section \ref{denoising-par} shows the results of the denoising technique using both procedures and the effect on the derived stellar parameter accuracy values. Finally, we conclude in Sec.~\ref{conclusion}.

\section{Datasets}
\label{database}
Two datasets were used in the context of the present study. The one analyzed in \cite{2023OAst...32..209G} and the one of \cite{2024Astro...3....1G}. The characteristics of these two datasets are described in Tab.~\ref{tab-databases}. The reason for selecting these diverse datasets is to check the procedure over different wavelength ranges and different resolving power. 

\begin{table}[!h]
    \centering
    \begin{tabular}{|c|c|c|}
    \hline
          Parameter  & Range  for DB1 & Range  for DB2 \\
          \hline
             \Teff & \multicolumn{2}{|c|}{3,600 -- 15,000 K} \\ 
        \logg & \multicolumn{2}{|c|}{2.0 -- 5.0 dex}\\
        \vsini & \multicolumn{2}{|c|}{0 -- 300 km/s} \\
        \met & \multicolumn{2}{|c|}{$-$1.5 -- 1.5 dex} \\
       $$ \micro$$ & \multicolumn{2}{|c|}{0 -- 4 km/s} \\ \hline
      Wavelength $\lambda$ & 4,450 -- 5,400 \AA & 8,400 -- 8,800 \AA \\
      Sampling in $\lambda$ & 0.05 \AA & 0.10 \AA \\
        Resolution ($\dfrac{\lambda}{\Delta \lambda}$)& 42,000 & 11,500 \\ \hline
    \end{tabular}
    \caption{Range of parameters used in the calculation of the synthetic spectra for the two datasets.}
    \label{tab-databases}
\end{table}

The steps of calculating the datasets are detailed in \cite{2022OAst...31...38G,2023OAst...32..209G} and \cite{2024Astro...3....1G}. In summary, line-blanketed model atmospheres are calculated using ATLAS9 \citep{Kurucz1992}. The models are plane parallel and in Local Thermodynamic Equilibrium (LTE). They are in hydrostatic and radiative equilibrium. We have calculated the models using the Opacity Distribution Function (ODF) of \cite{castelli}. Convection was included according to Smalley’s prescriptions \citep{2004IAUS..224..131S}. Convection is included in the atmospheres of stars cooler than 8,500 K using the mixing length theory. A mixing length parameter of 0.5 was used for 7,000 K $\leq$ Teff$\leq$ 8,500 K, and 1.25 for Teff$\leq$ 7000 K. 

We have used the radiative transfer code SYNSPEC \citep{2017arXiv170601859H} to calculate the synthetic spectra. As mentioned previously, two datasets were calculated with each one containing around 200,000 spectra. In both datasets, metal abundances were scaled with respect to the \cite{1998SSRv...85..161G} solar value from -1.5 dex up to +1.5 dex. The effective temperature, surface gravity, projected equatorial velocity, and microturbulence velocity were also modified according to the values displayed in Tab.~\ref{tab-databases}.  The first dataset consists of spectra having a resolution of 42,000 and a wavelength range between 4,450 and 5,400 \AA. As explained in \cite{2022OAst...31...38G, 2023OAst...32..209G}, this wavelength range is sensitive to all stellar parameters in the spectral range of AFGK stars. The second dataset has spectra computed between 8,400 and 8,800 \AA\ at a resolution of 11,500. This region includes the Gaia Radial Velocity Spectrometer (RVS, \citealt{2018A&A...616A...5C}). The RVS spectra contain lines sensitive to the stellar parameters and to the chemical abundance of many metals ( Mg, Si, Ca, Ti, Cr, Fe, Ni, and Zr, among others) at different ionization stages. The linelist used in this work is the one used in \cite{2022OAst...31...38G,2023OAst...32..209G}. It contains updated values for the atomic parameters such as the wavelength of the transitions, the oscillator strengths, the damping constants, and others.

\begin{figure*}
    \centering
    \includegraphics[width=0.48\linewidth]{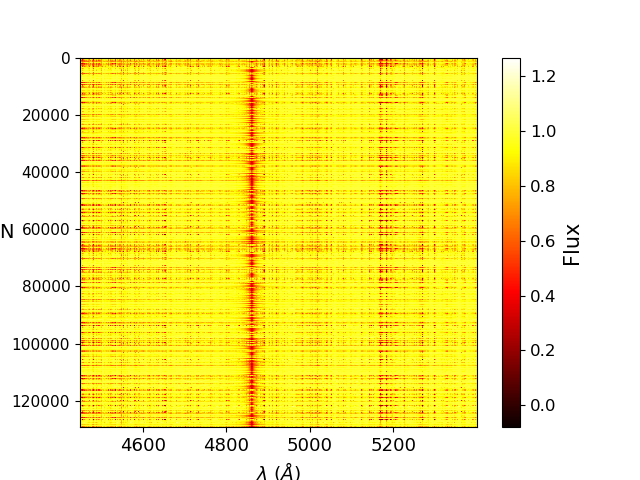}
    \includegraphics[width=0.48\linewidth]{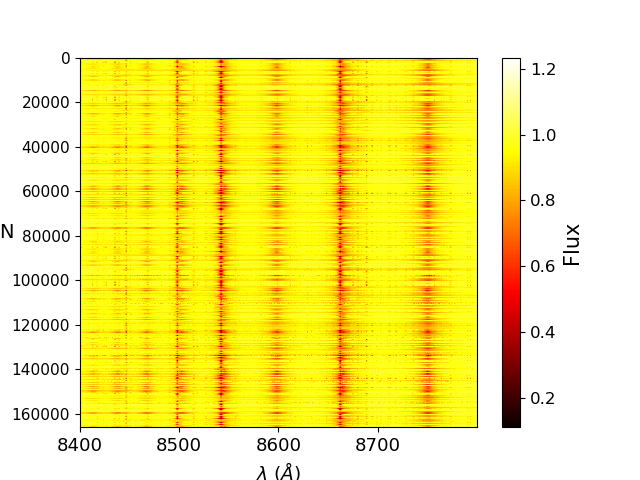}
    \caption{Color map representing the fluxes for a sample spectra of the two training datasets. The left color map represents dataset 1 and the right one represents dataset 2. The $y$-label represents the number of spectra, $N$.}
    \label{colormap}
\end{figure*}

In summary, we ended up with two datasets of around 200,000 synthetic spectra each, with \Teff, \logg, \vsini, \met, and \micro\ randomly chosen from Tab.~\ref{tab-databases}. Figure~\ref{colormap} shows a color map of a sub-sample of the datasets. The Balmer line is detected in the left color map for dataset 1 and the absorption lines of the calcium triplet ($\lambda$ = 8,498, 8,542, 8,662 \AA) are also shown in the color map of dataset 2 in the bottom part of the figure.

\begin{figure*}[!h]
    \centering
    \includegraphics[width=1.0\linewidth]{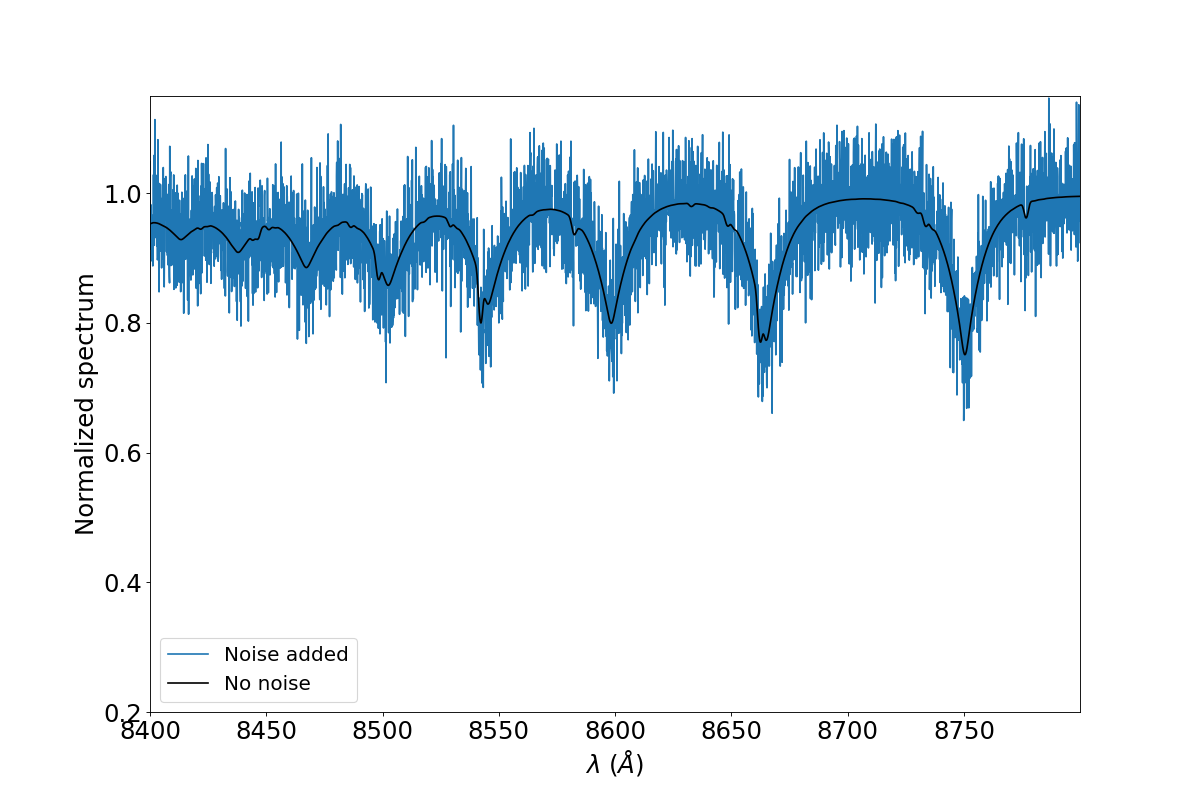}
    \caption{Example of a spectrum of dataset 2 calculated using a random selection of stellar parameters from Tab.\ref{tab-databases}. The black spectrum represents the synthetic spectrum calculated without noise and the blue one corresponds to the same parameters but with a SNR of 19. The stellar parameters of the spectrum are 14550 K, 3.05 dex, 44 \kms, -1.15 dex, and 3 \kms\ for \Teff, \logg, \vsini, \met, and \micro, respectively.}
    \label{fig:noise}
\end{figure*}

For each dataset, a set of spectra were calculated with random Gaussian noise between 5 and 300. This SNR is used to mimic the noisy observations that we will be denoising later on as they represent the average SNR encountered in real stellar spectra. An example of a spectrum calculated with and without noise in the parameter range of dataset 2 is shown in Fig.~\ref{fig:noise}.

\subsection{Data Augmentation}
We have also tested the effect of data augmentation in this work, and for that reason, we have calculated extra dataset as suggested in \cite{2022OAst...31...38G}. Data augmentation is a regularization technique that by increasing the diversity of the training data by applying different transformations to the existing one, helps in avoiding over-fitting and improves the predictions of stellar labels when applied with real observed data \citep{2023OAst...32..209G}. We have used the same approach of \cite{2022OAst...31...38G} in which 5 replicas of each spectrum in the dataset were performed. These replicas consist of 
\begin{itemize}
    \item Adding to each spectrum a Gaussian noise with a SNR ranging randomly between 5 and 300.
    \item The flux of each spectrum is multiplied with a scaling factor selected randomly between 0.95 and 1.05.
    \item The flux of each spectrum is multiplied with a new random scaling factor and noise was added.
    \item The flux of each spectrum is multiplied by a second-degree polynomial with values ranging between 0.95 and 1.05 and having its maximum randomly selected in the wavelength range of the dataset.
    \item The flux of each spectrum is multiplied by a second-degree polynomial and Gaussian noise added to it. 
\end{itemize}
For more details about data augmentation, we refer the reader to \cite{2022OAst...31...38G}.


\section{Auto-Encoders}
\label{AE}
Autoencoders, usually used in denoising and dimensionality reduction techniques \citep{37f2b6bee745402aa4e4d124d33be0e0, 0d5c5115ccbd4f9ebacc531e8ba0c706,1863696.1863746,Baldi2011AutoencodersUL,2014arXiv1404.7828S,2023A&A...677A.158E,2023MNRAS.526.3037S}, are a type of Neural Networks that work in an unsupervised way. They consist of two distinct yet similar algorithms, an encoder and a decoder. The encoder's role is to transform the spectra from a dimension of $N_{\lambda}$ flux point to a smaller size of $N_{\mathrm{latent}}$ inside a Latent Space. The decoder re-transform the $N_{\mathrm{latent}}$ to the original spectrum of $N_{\lambda}$ flux point. The choice of $N_{\mathrm{latent}}$ depends on the characteristics of the dataset. However, using the two datasets in this work, we found that the optimal size for the Latent Space is  $N_{\mathrm{latent}}=10$. This is found by minimizing the difference between the output spectra and the input one during the training process. It is true that different values of $N_{\mathrm{latent}}$ could be used, but our choice of $N_{\mathrm{latent}}$ was based on the smallest value that gives a reconstruction error less than 0.5\% as will be explained in the next steps. 

\begin{figure*}[!h]
    \centering
    \includegraphics[scale=0.6]{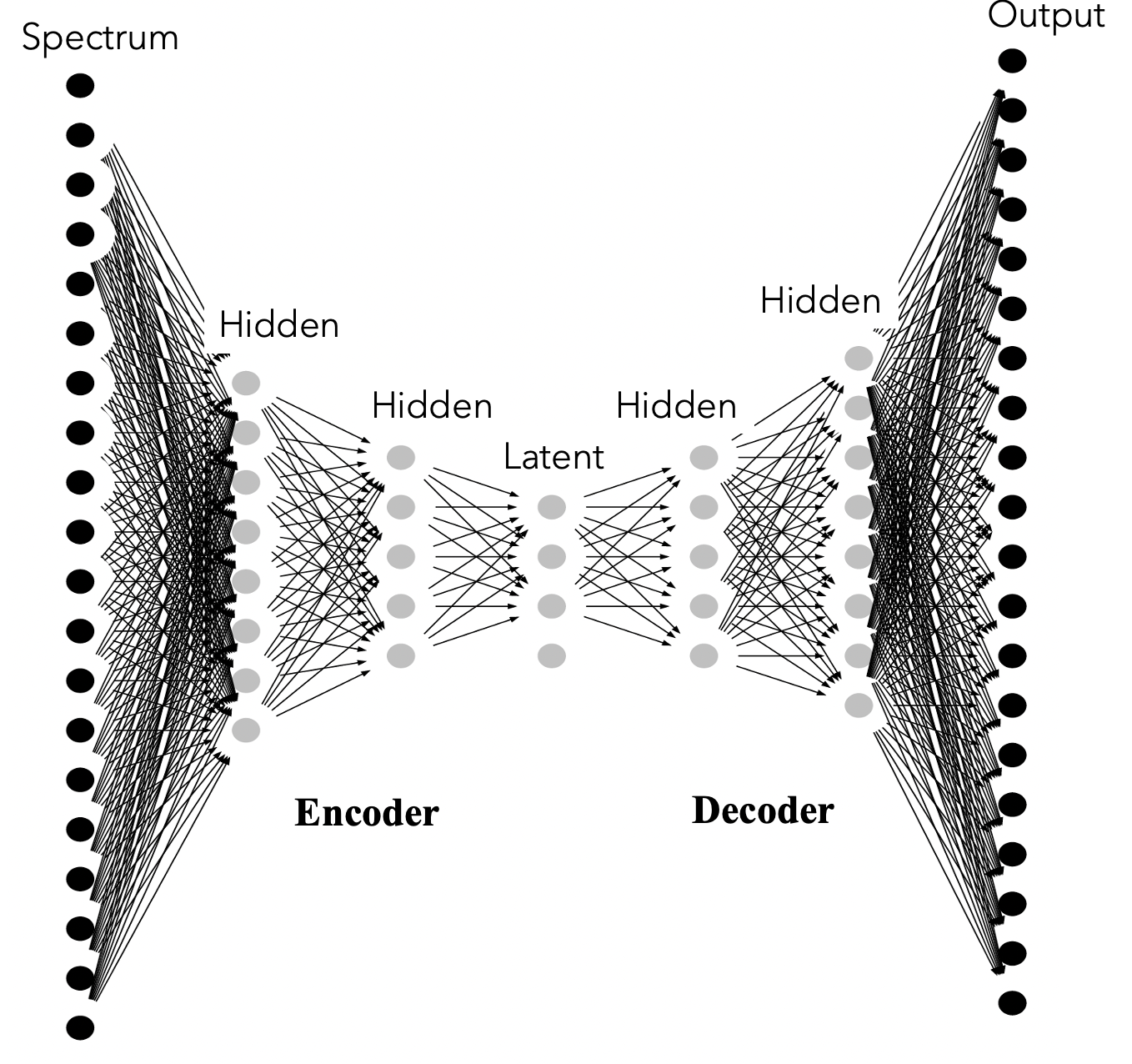}
    \caption{Sketch of an autoencoder that transforms an input spectrum of $N_{\lambda}$ data point to a lower dimension using a series of hidden layers (Encoder). The middle layer is the Latent Space. The  Decoder reconstructs the spectrum to its original dimension.}
    \label{autoencoder}
\end{figure*}

The classical architecture of an autoencoder is shown in Fig.~\ref{autoencoder} where the initial spectrum is introduced having 19,000 or 4,000 data points depending on the dataset and is then reduced to $N_{\mathrm{latent}}$ points through successive hidden layers. This first step defines the encoder part of the autoencoder. Then, the $N_{\mathrm{latent}}$ points are transformed to 19,000 or 4,000 data points while passing through different hidden layers. This second step defines the Decoder part of the autoencoder. The hidden layers are usually symmetrical in the encoder and decoder parts.

Two autoencoders were used in this work, one for each dataset. In both cases, the spectra are
reduced to 10 parameters in the Latent Space. The architecture of the used autoencoders is displayed in Tab.~\ref{auto-param}. We have used an Adam optimizer with a Mean Squared Error (MSE) loss function.

\begin{table*}[!thb]
\centering
\caption{Architecture of the autoencoder used for denoising.}
    \label{auto-param}
    \begin{tabular}{cc}
    \toprule
  Layer & Characteristics \\ \hline
    \multicolumn{2}{|c|}{Encoder}\\  \hline
     Input   & Spectrum of $N_{\lambda}$ data points  \\
     Hidden    &  1024 neurons \\
      Hidden    &  512 neurons  \\
     Hidden    &  256 neurons  \\
          Hidden    &  64 neurons  \\
     Hidden    &  32 neurons  \\ \hline \hline
    Latent Space    &  10 neurons \\ \hline \hline
    \multicolumn{2}{|c|}{Decoder}\\ \hline
Hidden    &  32 neurons  \\
Hidden    &  64 neurons  \\
 Hidden    &  256 neurons  \\
  Hidden    &  512 neurons  \\
  Hidden    &  1024 neurons \\
  Output  & Reconstructed spectrum of  $N_{\lambda}$   data points \\ 
  \bottomrule
    \end{tabular}
\end{table*}

Calculations were performed using \texttt{TensorFlow}\footnote{\url{https://www.tensorflow.org/}}  with the \texttt{Keras}\footnote{\url{https://keras.io/}} interface and were written in \texttt{Python}. 

The training of the autoencoders was performed using the 2 datasets containing the synthetic spectra with no noise. The convergence is achieved when the difference between the output and the input spectra is minimized through the MSE. Convergence usually occurs after around 500 epochs. For both datasets, we achieved an R$^2$ score larger than $0.995$. Meaning that the reconstruction of the spectra is performed with an error $<$0.5\%. Once the training is done, the denoising is performed when the trained autoencoders are applied to the noisy spectra.

\section{Principal Component Analysis}
\label{PCA}
PCA is a non-parametric mathematical transformation that extracts relevant information from a dataset \citep{WOLD198737,MACKIEWICZ1993303}. Its goal is to compute the most meaningful basis to represent a noisy dataset. expressanoisydataset. The new basis usually reveals hidden structure and filters out the noise \citep{2014arXiv1404.1100S}. PCA has been used for denoising \citep{BACCHELLI2006606,ZHANG20101531,dedde,8531438} or spectral dimension reduction \citep{MACKIEWICZ1993303,S4n,Gebran,2022OAst...31...38G,2023OAst...32..209G}. The main power of PCA is that it can reduce the dimension of the data while maintaining significant patterns and trends. 

The basic idea behind the use of PCA is to derive a small number of eigenvectors and use them to recover the information in the spectra. The steps of PCA calculation are
\begin{enumerate}
    \item The matrix containing the Training dataset has $N_{\lambda}$ flux points per spectrum, therefore the dataset can then be represented by a matrix $\textbf{\textit{M}}$ of size $N_{\mathrm{spectra}} \times N_{\lambda}$ where $N_{\mathrm{spectra}}$ represents the number of spectra in the dataset.\\
    \item The matrix $\textbf{\textit{M}}$ is then averaged along the $N_{\mathrm{spectra}}$-axis and this average is stored in a vector $\bar{M}$. \\
    \item The variance-covariance matrix $\textbf{\textit{C}}$  is calculated as 
\begin{equation}
\textbf{\textit{C}}=(\textbf{\textit{M}}-\bar{\textit{M}})^\mathrm{T}\cdot(\textbf{\textit{M}}-\bar{\textit{M}})\, 
\end{equation}
where the superscript "T" stands for the transpose operator. \\
\item The eigenvectors $\textbf{e}_k(\lambda)$ of $\textbf{\textit{C}}$ are then calculated. $\textbf{\textit{C}}$ has a dimension of $N_{\lambda}\times N_{\lambda}$. The Principal Components (PC) correspond to the eigenvectors sorted in decreasing magnitude. \\
\item Each spectrum of $\textbf{\textit{M}}$ is then projected on these PCs in order to find its corresponding coefficient $p_{jk}$ defined as 
\begin{equation}
p_{jk}=(M_j-\bar{M})\cdot \textbf{\textit{e}}_k
\end{equation}

\item  The original "denoised spectrum" can be calculated using
\begin{equation}
S_j=\bar{M}+\Sigma_{k=1}^{n_k} p_{jk}\textbf{\textit{e}}_k
 \end{equation}
\end{enumerate}

 The PCA can reduce the size of each spectrum from $N_{\lambda}$ to $n_k$. The choice of $n_k$ depends on the many parameters, the size of the dataset, the wavelength range, and the shape of the spectra lines. We have opted for a value for $n_k$ that reduces the mean reconstructed error to a value <0.5\% according to the following equation:

\begin{equation}
E(k_{max})=\left< \left( \dfrac{\vert\bar{M}+\Sigma_{k=1}^{n_k} p_{jk}\textbf{\textit{e}}_k - M_j\vert}{M_j}    \right)  \right>
\end{equation}

We have opted to a value for $n_k$ that reduces the mean reconstructed error to a value <0.5\%.  This value if found to be $n_k$=50. A detailed description of all steps of the PCA can be found in \cite{S4n, dms, Gebran, 2022OAst...31...38G, 2023OAst...32..209G, 2024Astro...3....1G}.  
For both datasets, we achieved an R$^2$ score larger than $0.996$.

\section{Denoising and parameters determination}
\label{denoising-par}
The datasets that contain the synthetic spectra without any added noise are used to train the autoencoder and to find the eigenvectors of the PCA procedure. These two techniques are then used on the set of noisy spectra that are calculated in Sec.~\ref{database}. The evaluation of the denoising procedure is tested in two ways. First, we checked the similarity of the denoised spectra with the original one with no noise added. Second, we checked the accuracy of the derived stellar parameters when we applied the procedures of \cite{2022OAst...31...38G,2023OAst...32..209G} on the denoised spectra from the autoencoder and PCA. 

\begin{figure*}[!h]
    \centering

    \includegraphics[scale=0.45]{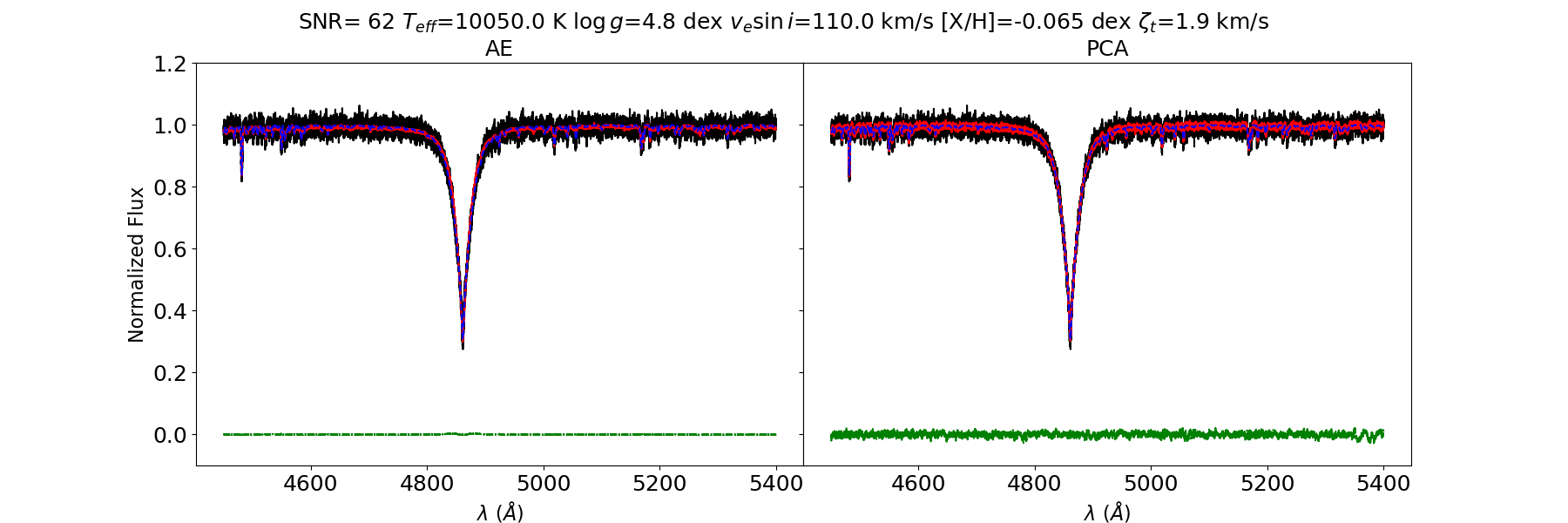}
    \includegraphics[scale=0.45]{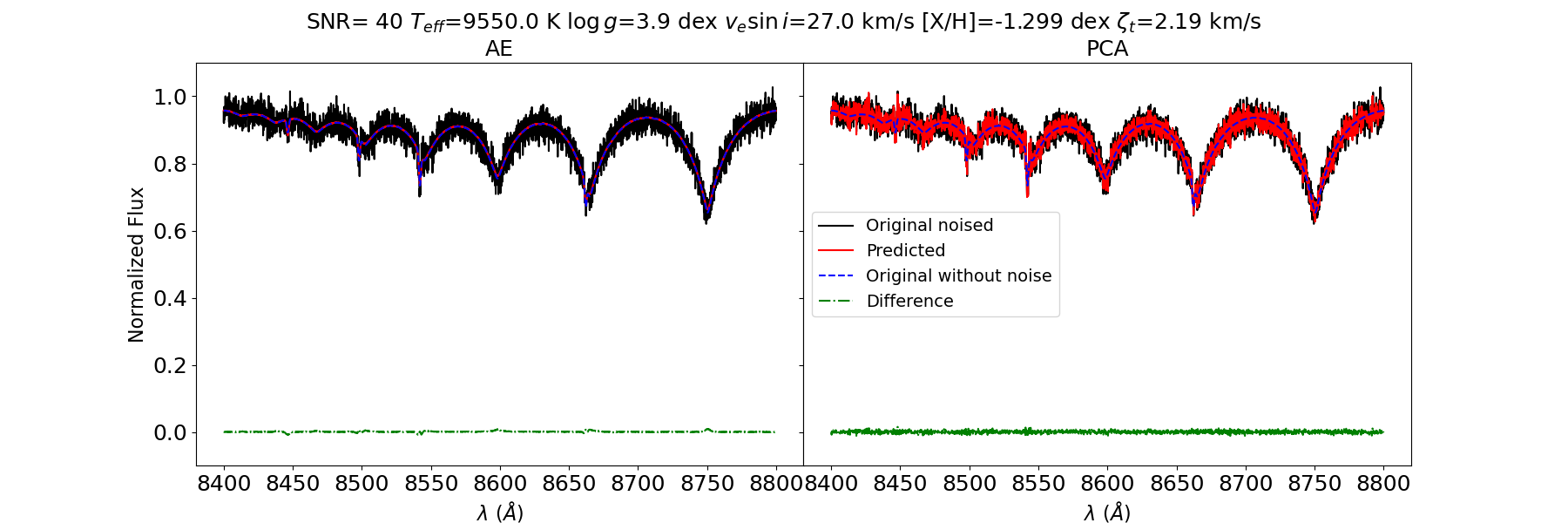}
    \caption{Denoising example of spectra having stellar parameters in both datasets. The upper plot displays a spectrum having the parameters of dataset 1 and the bottom one has the parameters of dataset 2. The noisy spectrum is in black, the original one without noise is in dashed blue, the denoised spectrum using the autoencoder (left panel) or PCA (right panel) technique is in red, and the difference between the denoised spectrum and the original one with no noise is in dash-dot green.}
    \label{fig:denoising}
\end{figure*}

Autoencoders usually replace PCA because of their non-linear properties, however, both techniques showed a good reconstruction power as shown by the R$^2$ score in Secs.~\ref{AE} and \ref{PCA}. A way to visualize the denoising of spectra is shown in Fig.~\ref{fig:denoising}. The figure is divided into two parts, the upper one displays a spectrum having the parameters of dataset 1 and the bottom one has the parameters of dataset 2. In each part, the noisy spectrum is in black, the original one without noise is in dashed blue, the denoised spectrum using the autoencoder (left panel) or PCA (right panel) technique is in red, and the difference between the denoised spectrum and the original one without noise is in dash-dot green. 

In \cite{2022OAst...31...38G,2023OAst...32..209G} we have introduced a technique to derive the stellar parameters of spectra using a Neural Network. We have used the same procedure to derive the accuracy of the stellar parameters once we apply the same technique to the denoised spectra. The main purpose of this step is not to evaluate if the derivation technique is accurate or not but it is to check how similar are the derived stellar parameters of the noisy spectra to the ones derived from the original spectra with no noise added.

The networks that we used are made of several fully dense layers and are trained to derive each parameters separately. The layers are described in Tab.~\ref{stellar-parameter-NN}. The first step of the analysis is to reduce the dimension of the spectra using a PCA procedure. This PCA is not related to the one used for denoising, it is just a step for optimizing the network and making the training faster (See \citealt{2022OAst...31...38G} for more details).

\begin{table}[!h]
\centering
        \caption{Architecture of the Neural Network used for stellar parameters determination.}
    \label{stellar-parameter-NN}
    \begin{tabular}{cc}
    \toprule  
    Layer & Characteristics \\ 
           \midrule
    Input   & Spectrum of $N_{\lambda}$ data points  \\
     PCA    &  Reduction to 50 data points  \\
      Hidden    &  4096 neurons  \\
     Hidden    &  2048 neurons  \\
          Hidden    &  1024 neurons \\
     Hidden    &  512 neurons  \\ 
Hidden    &  60 neurons  \\
  Output  & 1 parameter  \\ 
          \bottomrule
    \end{tabular}
\end{table}


Two different training are performed for each dataset. The first one is done using a dataset of only synthetic spectra with no noise added and the second one consists of applying data augmentation with spectra having a range of SNR between 3 and 300. 

Because we already know the stellar parameters of the spectra, the evaluation is performed by calculating the difference between the predicted parameter and the original one using the equation 
\begin{equation}
    \text{Accuracy}=\frac{1}{N}\sqrt{\sum_{i=1}^{N} (\text{Predicted}-\text{Original})^2}
\end{equation}
where $N$ is the total number of noisy spectra used in the evaluation. This is done for \Teff, \logg, \vsini, \micro, and \met. Tables \ref{tab:afgk} and \ref{tab:gaia} display the accuracy values for the parameters for the two datasets when deriving the stellar labels of $\sim$25,000 with no noise added (column 2), with random noise (column 3), with random noise then denoised using autoencoder of Sec.~\ref{AE} (column 4) and using PCA of Sec.~\ref{PCA} (column 5). Each table is divided into two, one part when data augmentation is performed and one without it.

\begin{table*}[!h]
    \caption{Accuracy values on the derived stellar parameters for the spectra calculated using the parameters of dataset 1. The spectra are calculated with no noise added (Col. 2), with random Gaussian noise (Col. 3), with random noise and then denoised using the autoencoder network (Col. 4), and denoised using PCA (Col. 5). }
    \label{tab:afgk}

		\begin{tabular}{ccccc}
    \toprule
    \multicolumn{5}{|c|}{Augmented dataset} \\ \hline 
         Parameters &   No noise & With noise & Denoised (AE) & Denoised (PCA) \\
         \Teff\ (K) & 52 & 181 & 240 & 190\\
         \logg\ (dex) & 0.017 & 0.089 & 0.160 & 0.100 \\
         \vsini\ (\kms) & 1.80 & 7.58 & 9.74 &  7.60\\
         $\xi_t$ (\kms)& 0.09 & 0.22 & 0.32 & 0.23\\
         \met\ (dex) & 0.021 & 0.071 & 0.103 & 0.071\\ 
         \hline
    \multicolumn{5}{|c|}{No data Augmentation} \\ \hline 
         Parameters &   No noise & With noise & Denoised (AE) & Denoised (PCA) \\
         \Teff\ (K) & 47 & 219 & 243 & 222\\
         \logg\ (dex) & 0.018 & 0.121 & 0.168 & 0.121 \\
         \vsini\ (\kms) & 1.69 & 9.49 & 10.10 & 9.50\\
         $\xi_t$ (\kms)& 0.11& 0.40 & 0.40 & 0.48 \\
         \met\ (dex) &0.019 & 0.097 & 0.106 &0.098 \\ 
                 \bottomrule
    \end{tabular}
\end{table*}

\begin{table*}[!h]
    \caption{Same as Tab.~\ref{tab:afgk} for the parameters in dataset 2.}
    \label{tab:gaia}
		\begin{tabular}{ccccc}
    \toprule
    \multicolumn{5}{|c|}{Augmented dataset} \\ \hline 
         Parameters &   No noise & With noise & Denoised (AE) & Denoised (PCA) \\
         \Teff\ (K) &  116 & 300 & 303& 345 \\
         \logg\ (dex) &  0.037 & 0.145 & 0.174 & 0.176 \\
         \vsini\ (\kms) &  6.61 & 13.40 & 16.16 & 13.87 \\
         $\xi_t$ (\kms)&  0.16 & 0.63 & 0.79 & 0.63\\
         \met\ (dex) & 0.038 & 0.170 & 0.204 &  0.172\\ 
         \hline
    \multicolumn{5}{|c|}{No data Augmentation} \\ \hline 
         Parameters &   No noise & With noise & Denoised (AE) & Denoised (PCA) \\
         \Teff\ (K) &124 & 360 & 576 & 577 \\
         \logg\ (dex) &   0.031 & 0.188 & 0.310 & 0.310 \\
         \vsini\ (\kms) & 5.93 & 17.50 & 17.60& 21.70 \\
         $\xi_t$ (\kms)& 0.09 & 1.12 & 1.36&  1.36 \\
         \met\ (dex) & 0.035 & 0.262 & 0.272 & 0.267\\ 
                   \bottomrule          
    \end{tabular}
\end{table*}

\begin{figure*}[!h]
    \centering

    \includegraphics[scale=0.45]{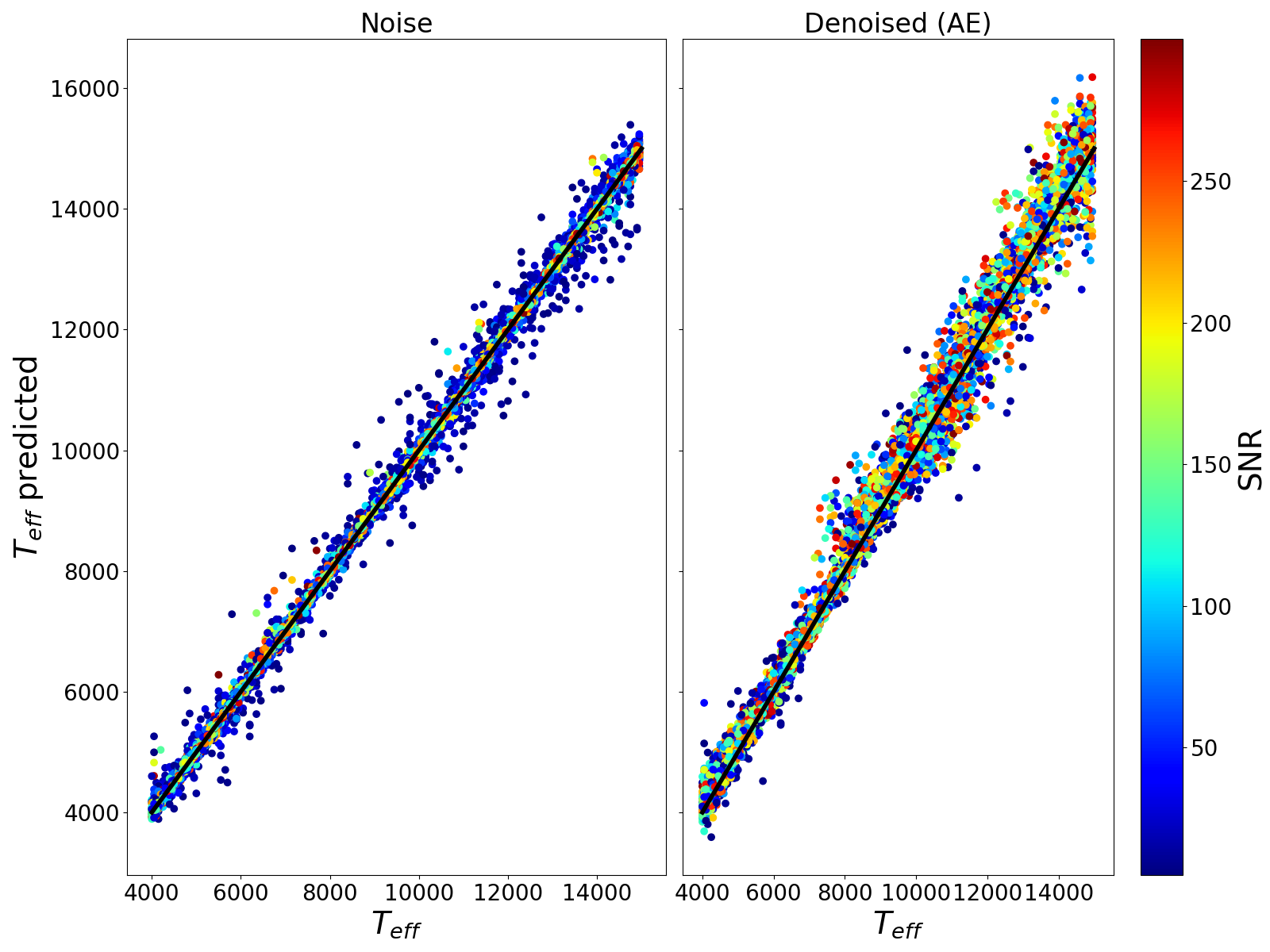}
    \caption{Predicted \Teff\ as a function of the true values for the data with noise (left panel) and the denoised data using autoencoder (right panel). The spectra are from the augmented dataset 2. The data are color-coded to the SNR values. }
    \label{fig:merit}
\end{figure*}

A detailed analysis of Tabs.~\ref{tab:afgk} and \ref{tab:gaia} show that:
\begin{itemize}
    \item Data augmentation is an important step to be applied if we need to derive the stellar parameters of noisy spectra. Without it, the model will only learn to derive the parameters of synthetic spectra without any noise added. A similar conclusion was also found in \cite{2023OAst...32..209G}.\\
    \item PCA denoising is capable of recovering the line profile and the details in the spectra. This is reflected by comparing the accuracy values of the derived parameters using the denoised spectra from the autoencoders and PCA (i.e. comparing Cols. 4 and 5).\\
    \item The parameters derived using the PCA denoising technique are more accurate than the ones derived using the autoencoder denoising.\\
    \item No denoising technique is capable of improving the accuracy of the stellar parameters for the one directly derived from noisy spectra (displayed in Col. 3). \\
    \item The stellar parameter algorithm is capable of deriving the stellar labels without the need for a denoising technique. \\
\end{itemize}
 These tests show mainly that data augmentation is very important when Neural Networks are used to derive the stellar parameters of noisy spectra, a results already found by \cite{2022OAst...31...38G, 2023OAst...32..209G}. As an example, Fig. \ref{fig:merit} displays the predicted \Teff\ with respect to the original one for the data with noise from the augmented dataset 2 (left panel) and the denoised data using autoencoder (right panel) from the same dataset. The data are color-coded to the SNR values. The straight black line represents the best prediction line ($x=y$). The left panel shows that the highly dispersed results are the ones for the low SNR spectra. Once the spectra are denoised, the dispersion appears to be present for all SNR values with no specific trend or deviation. This is true for all stellar parameters.  Independently of the denoising technique, there is no improvement found in the accuracy values of the derived parameters of denoised spectra when the networks were trained on noisy spectra. Applying the networks on noisy data gives more accurate results then when it is applied on denoised data.

\section{Conclusion}
\label{conclusion}
In this work, we have applied two different denoising techniques, an autoencoder, and a PCA, on spectra with random Gaussian noise added to derive the stellar parameters using Neural Networks of \cite{2022OAst...31...38G, 2023OAst...32..209G}. The method was applied to two different spectra ranges, one in 4,450--5,400 \AA\ and one in the Gaia RVS range from 8,400--8,800 \AA. In this study, we do not constrain the stellar parameter derivation technique, this was done previously in \cite{2022OAst...31...38G,2023OAst...32..209G}.
Interestingly, when applying the model to denoised spectra, there was no noticeable improvement in the accuracy of the derived fundamental parameters, such as \Teff, \logg, \vsini, \micro, and \met. This outcome was unexpected, as denoising is typically thought to enhance the precision of predictions. However, the results indicate that data augmentation plays a more crucial role. When the model is trained on datasets that include noise, the accuracy of predictions for noisy spectra improves significantly, suggesting that the network becomes better equipped to handle real observed spectra. This highlights the importance of incorporating noisy data into training rather than relying on post-processing techniques like denoising to improve accuracy.
To further validate these findings, it would be valuable to explore other denoising techniques and assess their impact on prediction accuracy. Techniques such as those presented in \cite{Alsberg1997WaveletDO}, \cite{2018NatSR...814351K}, and \cite{ZHAO2021119374} could be tested to see if they yield better results in reducing noise while maintaining or enhancing the precision of derived parameters. These additional experiments would help solidify the conclusion that data augmentation is more effective than denoising in improving the accuracy of noisy spectra predictions, offering deeper insights into how best to model real observational spectra.

\textbf{Acknowledgment}: The authors acknowledge Saint Mary's College for providing the high-power computing cluster used in this work. The authors are grateful for the reviewer’s valuable comments that improved the manuscript.  \\

\textbf{Funding information}: Authors state no funding involved.  \\

\textbf{Author contributions}: All authors have accepted responsibility for the entire content of this manuscript and consented to its submission to the journal, reviewed all the results, and approved the final version of the manuscript. MG and RB designed the code and carried out the calculations. MG prepared the manuscript with contributions from all co-authors.  \\

\textbf{Conflict of interest}: The authors state no conflict of interest.

\bibliographystyle{apalike}
\bibliography{biblio}

\begin{thebibliography}{}

\bibitem[Alsberg et~al., 1997]{Alsberg1997WaveletDO}
Alsberg, B.~K., Woodward, A.~M., Winson, M.~K., Rowland, J.~J., and Kell, D.~B.
  (1997).
\newblock Wavelet denoising of infrared spectra.
\newblock {\em Analyst}, 122:645--652.

\bibitem[Bacchelli and Papi, 2006]{BACCHELLI2006606}
Bacchelli, S. and Papi, S. (2006).
\newblock Image denoising using principal component analysis in the wavelet
  domain.
\newblock {\em Journal of Computational and Applied Mathematics},
  189(1):606--621.
\newblock Proceedings of The 11th International Congress on Computational and
  Applied Mathematics.

\bibitem[Baldi, 2011]{Baldi2011AutoencodersUL}
Baldi, P. (2011).
\newblock Autoencoders, unsupervised learning, and deep architectures.
\newblock In {\em ICML Unsupervised and Transfer Learning}.

\bibitem[Ballard, 1987]{1863696.1863746}
Ballard, D.~H. (1987).
\newblock Modular learning in neural networks.
\newblock In {\em Proceedings of the Sixth National Conference on Artificial
  Intelligence - Volume 1}, AAAI'87, page 279–284. AAAI Press.

\bibitem[{Castelli} and {Kurucz}, 2003]{castelli}
{Castelli}, F. and {Kurucz}, R.~L. (2003).
\newblock {New Grids of ATLAS9 Model Atmospheres}.
\newblock In {Piskunov}, N., {Weiss}, W.~W., and {Gray}, D.~F., editors, {\em
  Modelling of Stellar Atmospheres}, volume 210, page A20.

\bibitem[{Chung}, 2020]{2020arXiv200709539C}
{Chung}, M.~K. (2020).
\newblock {Gaussian kernel smoothing}.
\newblock {\em arXiv e-prints}, page arXiv:2007.09539.

\bibitem[{Cropper} et~al., 2018]{2018A&A...616A...5C}
{Cropper}, M., {Katz}, D., {Sartoretti}, P., {Prusti}, T., {de Bruijne},
  J.~H.~J., {Chassat}, F., {Charvet}, P., {Boyadjian}, J., {Perryman}, M.,
  {Sarri}, G., {Gare}, P., {Erdmann}, M., {Munari}, U., {Zwitter}, T.,
  {Wilkinson}, M., {Arenou}, F., {Vallenari}, A., {G{\'o}mez}, A., {Panuzzo},
  P., {Seabroke}, G., {Allende Prieto}, C., {Benson}, K., {Marchal}, O.,
  {Huckle}, H., {Smith}, M., {Dolding}, C., {Jan{\ss}en}, K., {Viala}, Y.,
  {Blomme}, R., {Baker}, S., {Boudreault}, S., {Crifo}, F., {Soubiran}, C.,
  {Fr{\'e}mat}, Y., {Jasniewicz}, G., {Guerrier}, A., {Guy}, L.~P., {Turon},
  C., {Jean-Antoine-Piccolo}, A., {Th{\'e}venin}, F., {David}, M., {Gosset},
  E., and {Damerdji}, Y. (2018).
\newblock {Gaia Data Release 2. Gaia Radial Velocity Spectrometer}.
\newblock {\em \aap}, 616:A5.

\bibitem[{Einig} et~al., 2023]{2023A&A...677A.158E}
{Einig}, L., {Pety}, J., {Roueff}, A., {Vandame}, P., {Chanussot}, J., {Gerin},
  M., {Orkisz}, J.~H., {Palud}, P., {Santa-Maria}, M.~G., {de Souza Magalhaes},
  V., and et~al. (2023).
\newblock {Deep learning denoising by dimension reduction: Application to the
  ORION-B line cubes}.
\newblock {\em \aap}, 677:A158.

\bibitem[{Fogelman Soulie} et~al., 1987]{0d5c5115ccbd4f9ebacc531e8ba0c706}
{Fogelman Soulie}, F., Gallinari, P., Lecun, Y., and Thiria, S. (1987).
\newblock {\em Automata networks and artificial intelligence}, pages 133--186.
\newblock Princeton University Press.

\bibitem[{Gebran}, 2024]{2024Astro...3....1G}
{Gebran}, M. (2024).
\newblock {Generating Stellar Spectra Using Neural Networks}.
\newblock {\em Astronomy}, 3(1):1--13.

\bibitem[{Gebran} et~al., 2022]{2022OAst...31...38G}
{Gebran}, M., {Connick}, K., {Farhat}, H., {Paletou}, F., and {Bentley}, I.
  (2022).
\newblock {Deep learning application for stellar parameters determination:
  I-constraining the hyperparameters}.
\newblock {\em Open Astronomy}, 31(1):38--57.

\bibitem[{Gebran} et~al., 2016]{Gebran}
{Gebran}, M., {Farah}, W., {Paletou}, F., {Monier}, R., and {Watson}, V.
  (2016).
\newblock {A new method for the inversion of atmospheric parameters of A/Am
  stars}.
\newblock {\em \aap}, 589:A83.

\bibitem[{Gebran} et~al., 2023]{2023OAst...32..209G}
{Gebran}, M., {Paletou}, F., {Bentley}, I., {Brienza}, R., and {Connick}, K.
  (2023).
\newblock {Deep learning applications for stellar parameter determination:
  II-application to the observed spectra of AFGK stars}.
\newblock {\em Open Astronomy}, 32(1):209.

\bibitem[{Gheller} and {Vazza}, 2022]{2022MNRAS.509..990G}
{Gheller}, C. and {Vazza}, F. (2022).
\newblock {Convolutional deep denoising autoencoders for radio astronomical
  images}.
\newblock {\em \mnras}, 509(1):990--1009.

\bibitem[{Grevesse} and {Sauval}, 1998]{1998SSRv...85..161G}
{Grevesse}, N. and {Sauval}, A.~J. (1998).
\newblock {Standard Solar Composition}.
\newblock {\em \ssr}, 85:161--174.

\bibitem[Halidou et~al., 2023]{wavelet}
Halidou, A., Mohamadou, Y., Ari, A. A.~A., and Zacko, E. J.~G. (2023).
\newblock Review of wavelet denoising algorithms.
\newblock {\em Multimedia Tools Appl.}, 82(27):41539–41569.

\bibitem[{Hubeny} and {Lanz}, 2017]{2017arXiv170601859H}
{Hubeny}, I. and {Lanz}, T. (2017).
\newblock {A brief introductory guide to TLUSTY and SYNSPEC}.
\newblock {\em arXiv e-prints}, page arXiv:1706.01859.

\bibitem[Jonas, 2009]{5109671}
Jonas, J.~L. (2009).
\newblock Meerkat—the south african array with composite dishes and wide-band
  single pixel feeds.
\newblock {\em Proceedings of the IEEE}, 97(8):1522--1530.

\bibitem[{Koziol} et~al., 2018]{2018NatSR...814351K}
{Koziol}, P., {Raczkowska}, M.~K., {Skibinska}, J., {Urbaniak-Wasik}, S.,
  {Paluszkiewicz}, C., {Kwiatek}, W., and {Wrobel}, T.~P. (2018).
\newblock {Comparison of spectral and spatial denoising techniques in the
  context of High Definition FT-IR imaging hyperspectral data}.
\newblock {\em Scientific Reports}, 8:14351.

\bibitem[Kumar and Sodhi, 2020]{9083712}
Kumar, A. and Sodhi, S.~S. (2020).
\newblock Comparative analysis of gaussian filter, median filter and denoise
  autoenocoder.
\newblock In {\em 2020 7th International Conference on Computing for
  Sustainable Global Development (INDIACom)}, pages 45--51.

\bibitem[{Kurucz}, 1992]{Kurucz1992}
{Kurucz}, R.~L. (1992).
\newblock {Atomic and Molecular Data for Opacity Calculations}.
\newblock {\em \rmxaa}, 23:45.

\bibitem[Lecun, 1987]{37f2b6bee745402aa4e4d124d33be0e0}
Lecun, Y. (1987).
\newblock {\em PhD thesis: Modeles connexionnistes de l'apprentissage
  (connectionist learning models)}.
\newblock Universite P. et M. Curie (Paris 6).

\bibitem[Li, 2018]{8531438}
Li, B. (2018).
\newblock A principal component analysis approach to noise removal for speech
  denoising.
\newblock In {\em 2018 International Conference on Virtual Reality and
  Intelligent Systems (ICVRIS)}, pages 429--432.

\bibitem[Maćkiewicz and Ratajczak, 1993]{MACKIEWICZ1993303}
Maćkiewicz, A. and Ratajczak, W. (1993).
\newblock Principal components analysis (pca).
\newblock {\em Computers \& Geosciences}, 19(3):303--342.

\bibitem[Murali et~al., 2012]{dedde}
Murali, Y., Babu, M., Subramanyam, M., and Prasad, D. (2012).
\newblock Pca based image denoising.
\newblock {\em Signal \& Image Processing}, 3.

\bibitem[{Offringa} et~al., 2013]{2013A&A...549A..11O}
{Offringa}, A.~R., {de Bruyn}, A.~G., {Zaroubi}, S., {van Diepen}, G.,
  {Martinez-Ruby}, O., {Labropoulos}, P., {Brentjens}, M.~A., {Ciardi}, B.,
  {Daiboo}, S., {Harker}, G., {Jeli{\'c}}, V., {Kazemi}, S., {Koopmans},
  L.~V.~E., {Mellema}, G., {Pandey}, V.~N., {Pizzo}, R.~F., {Schaye}, J.,
  {Vedantham}, H., {Veligatla}, V., {Wijnholds}, S.~J., {Yatawatta}, S.,
  {Zarka}, P., {Alexov}, A., {Anderson}, J., {Asgekar}, A., {Avruch}, M.,
  {Beck}, R., {Bell}, M., {Bell}, M.~R., {Bentum}, M., {Bernardi}, G., {Best},
  P., {Birzan}, L., {Bonafede}, A., {Breitling}, F., {Broderick}, J.~W.,
  {Br{\"u}ggen}, M., {Butcher}, H., {Conway}, J., {de Vos}, M., {Dettmar},
  R.~J., {Eisloeffel}, J., {Falcke}, H., {Fender}, R., {Frieswijk}, W.,
  {Gerbers}, M., {Griessmeier}, J.~M., {Gunst}, A.~W., {Hassall}, T.~E.,
  {Heald}, G., {Hessels}, J., {Hoeft}, M., {Horneffer}, A., {Karastergiou}, A.,
  {Kondratiev}, V., {Koopman}, Y., {Kuniyoshi}, M., {Kuper}, G., {Maat}, P.,
  {Mann}, G., {McKean}, J., {Meulman}, H., {Mevius}, M., {Mol}, J.~D.,
  {Nijboer}, R., {Noordam}, J., {Norden}, M., {Paas}, H., {Pandey}, M.,
  {Pizzo}, R., {Polatidis}, A., {Rafferty}, D., {Rawlings}, S., {Reich}, W.,
  {R{\"o}ttgering}, H.~J.~A., {Schoenmakers}, A.~P., {Sluman}, J., {Smirnov},
  O., {Sobey}, C., {Stappers}, B., {Steinmetz}, M., {Swinbank}, J., {Tagger},
  M., {Tang}, Y., {Tasse}, C., {van Ardenne}, A., {van Cappellen}, W., {van
  Duin}, A.~P., {van Haarlem}, M., {van Leeuwen}, J., {van Weeren}, R.~J.,
  {Vermeulen}, R., {Vocks}, C., {Wijers}, R.~A.~M.~J., {Wise}, M., and
  {Wucknitz}, O. (2013).
\newblock {The LOFAR radio environment}.
\newblock {\em \aap}, 549:A11.

\bibitem[{Paletou} et~al., 2015a]{S4n}
{Paletou}, F., {B{\"o}hm}, T., {Watson}, V., and {Trouilhet}, J.~F. (2015a).
\newblock {Inversion of stellar fundamental parameters from ESPaDOnS and Narval
  high-resolution spectra}.
\newblock {\em \aap}, 573:A67.

\bibitem[{Paletou} et~al., 2015b]{dms}
{Paletou}, F., {Gebran}, M., {Houdebine}, E.~R., and {Watson}, V. (2015b).
\newblock {Principal component analysis-based inversion of effective
  temperatures for late-type stars}.
\newblock {\em \aap}, 580:A78.

\bibitem[{Schmidhuber}, 2014]{2014arXiv1404.7828S}
{Schmidhuber}, J. (2014).
\newblock {Deep Learning in Neural Networks: An Overview}.
\newblock {\em arXiv e-prints}, page arXiv:1404.7828.

\bibitem[{Scourfield} et~al., 2023]{2023MNRAS.526.3037S}
{Scourfield}, M., {Saintonge}, A., {de Mijolla}, D., and {Viti}, S. (2023).
\newblock {De-noising of galaxy optical spectra with autoencoders}.
\newblock {\em \mnras}, 526(2):3037--3050.

\bibitem[{Shlens}, 2014]{2014arXiv1404.1100S}
{Shlens}, J. (2014).
\newblock {A Tutorial on Principal Component Analysis}.
\newblock {\em arXiv e-prints}, page arXiv:1404.1100.

\bibitem[{Smalley}, 2004]{2004IAUS..224..131S}
{Smalley}, B. (2004).
\newblock {Observations of convection in A-type stars}.
\newblock In {Zverko}, J., {Ziznovsky}, J., {Adelman}, S.~J., and {Weiss},
  W.~W., editors, {\em The A-Star Puzzle}, volume 224, pages 131--138.

\bibitem[{Tingay} et~al., 2013]{2013PASA...30....7T}
{Tingay}, S.~J., {Goeke}, R., {Bowman}, J.~D., {Emrich}, D., {Ord}, S.~M.,
  {Mitchell}, D.~A., {Morales}, M.~F., {Booler}, T., {Crosse}, B., {Wayth},
  R.~B., {Lonsdale}, C.~J., {Tremblay}, S., {Pallot}, D., {Colegate}, T.,
  {Wicenec}, A., {Kudryavtseva}, N., {Arcus}, W., {Barnes}, D., {Bernardi}, G.,
  {Briggs}, F., {Burns}, S., {Bunton}, J.~D., {Cappallo}, R.~J., {Corey},
  B.~E., {Deshpande}, A., {Desouza}, L., {Gaensler}, B.~M., {Greenhill}, L.~J.,
  {Hall}, P.~J., {Hazelton}, B.~J., {Herne}, D., {Hewitt}, J.~N.,
  {Johnston-Hollitt}, M., {Kaplan}, D.~L., {Kasper}, J.~C., {Kincaid}, B.~B.,
  {Koenig}, R., {Kratzenberg}, E., {Lynch}, M.~J., {Mckinley}, B., {Mcwhirter},
  S.~R., {Morgan}, E., {Oberoi}, D., {Pathikulangara}, J., {Prabu}, T.,
  {Remillard}, R.~A., {Rogers}, A.~E.~E., {Roshi}, A., {Salah}, J.~E., {Sault},
  R.~J., {Udaya-Shankar}, N., {Schlagenhaufer}, F., {Srivani}, K.~S.,
  {Stevens}, J., {Subrahmanyan}, R., {Waterson}, M., {Webster}, R.~L.,
  {Whitney}, A.~R., {Williams}, A., {Williams}, C.~L., and {Wyithe}, J.~S.~B.
  (2013).
\newblock {The Murchison Widefield Array: The Square Kilometre Array Precursor
  at Low Radio Frequencies}.
\newblock {\em \pasa}, 30:e007.

\bibitem[Wold et~al., 1987]{WOLD198737}
Wold, S., Esbensen, K., and Geladi, P. (1987).
\newblock Principal component analysis.
\newblock {\em Chemometrics and Intelligent Laboratory Systems}, 2(1):37--52.
\newblock Proceedings of the Multivariate Statistical Workshop for Geologists
  and Geochemists.

\bibitem[{York} et~al., 2000]{2000AJ....120.1579Y}
{York}, D.~G., {Adelman}, J., {Anderson}, John~E., J., {Anderson}, S.~F.,
  {Annis}, J., {Bahcall}, N.~A., {Bakken}, J.~A., {Barkhouser}, R., {Bastian},
  S., {Berman}, E., {Boroski}, W.~N., {Bracker}, S., {Briegel}, C., {Briggs},
  J.~W., {Brinkmann}, J., {Brunner}, R., {Burles}, S., {Carey}, L., {Carr},
  M.~A., {Castander}, F.~J., {Chen}, B., {Colestock}, P.~L., {Connolly}, A.~J.,
  {Crocker}, J.~H., {Csabai}, I., {Czarapata}, P.~C., {Davis}, J.~E., {Doi},
  M., {Dombeck}, T., {Eisenstein}, D., {Ellman}, N., {Elms}, B.~R., {Evans},
  M.~L., {Fan}, X., {Federwitz}, G.~R., {Fiscelli}, L., {Friedman}, S.,
  {Frieman}, J.~A., {Fukugita}, M., {Gillespie}, B., {Gunn}, J.~E., {Gurbani},
  V.~K., {de Haas}, E., {Haldeman}, M., {Harris}, F.~H., {Hayes}, J.,
  {Heckman}, T.~M., {Hennessy}, G.~S., {Hindsley}, R.~B., {Holm}, S.,
  {Holmgren}, D.~J., {Huang}, C.-h., {Hull}, C., {Husby}, D., {Ichikawa},
  S.-I., {Ichikawa}, T., {Ivezi{\'c}}, {\v{Z}}., {Kent}, S., {Kim}, R. S.~J.,
  {Kinney}, E., {Klaene}, M., {Kleinman}, A.~N., {Kleinman}, S., {Knapp},
  G.~R., {Korienek}, J., {Kron}, R.~G., {Kunszt}, P.~Z., {Lamb}, D.~Q., {Lee},
  B., {Leger}, R.~F., {Limmongkol}, S., {Lindenmeyer}, C., {Long}, D.~C.,
  {Loomis}, C., {Loveday}, J., {Lucinio}, R., {Lupton}, R.~H., {MacKinnon}, B.,
  {Mannery}, E.~J., {Mantsch}, P.~M., {Margon}, B., {McGehee}, P., {McKay},
  T.~A., {Meiksin}, A., {Merelli}, A., {Monet}, D.~G., {Munn}, J.~A.,
  {Narayanan}, V.~K., {Nash}, T., {Neilsen}, E., {Neswold}, R., {Newberg},
  H.~J., {Nichol}, R.~C., {Nicinski}, T., {Nonino}, M., {Okada}, N., {Okamura},
  S., {Ostriker}, J.~P., {Owen}, R., {Pauls}, A.~G., {Peoples}, J., {Peterson},
  R.~L., {Petravick}, D., {Pier}, J.~R., {Pope}, A., {Pordes}, R., {Prosapio},
  A., {Rechenmacher}, R., {Quinn}, T.~R., {Richards}, G.~T., {Richmond}, M.~W.,
  {Rivetta}, C.~H., {Rockosi}, C.~M., {Ruthmansdorfer}, K., {Sandford}, D.,
  {Schlegel}, D.~J., {Schneider}, D.~P., {Sekiguchi}, M., {Sergey}, G.,
  {Shimasaku}, K., {Siegmund}, W.~A., {Smee}, S., {Smith}, J.~A., {Snedden},
  S., {Stone}, R., {Stoughton}, C., {Strauss}, M.~A., {Stubbs}, C., {SubbaRao},
  M., {Szalay}, A.~S., {Szapudi}, I., {Szokoly}, G.~P., {Thakar}, A.~R.,
  {Tremonti}, C., {Tucker}, D.~L., {Uomoto}, A., {Vanden Berk}, D., {Vogeley},
  M.~S., {Waddell}, P., {Wang}, S.-i., {Watanabe}, M., {Weinberg}, D.~H.,
  {Yanny}, B., {Yasuda}, N., and {SDSS Collaboration} (2000).
\newblock {The Sloan Digital Sky Survey: Technical Summary}.
\newblock {\em \aj}, 120(3):1579--1587.

\bibitem[Zhang et~al., 2010]{ZHANG20101531}
Zhang, L., Dong, W., Zhang, D., and Shi, G. (2010).
\newblock Two-stage image denoising by principal component analysis with local
  pixel grouping.
\newblock {\em Pattern Recognition}, 43(4):1531--1549.

\bibitem[Zhao et~al., 2021]{ZHAO2021119374}
Zhao, X., Liu, G., Sui, Y., Xu, M., and Tong, L. (2021).
\newblock Denoising method for raman spectra with low signal-to-noise ratio
  based on feature extraction.
\newblock {\em Spectrochimica Acta Part A: Molecular and Biomolecular
  Spectroscopy}, 250:119374.

\end{thebibliography}

\end{document}